# Personal computer realizations for two classics of quantum chemistry

S. M. Blinder
University of Michigan
Ann Arbor, MI 48109-1055
email: sblinder@umich.edu

## Abstract

Hylleraas in 1929 carried out a variational computation on the Schrödinger equation for the helium atom which gave, for the first time, a ground-state energy in essential agreement with experimental results. Coolidge and James in 1933, likewise did the first accurate computation for the hydrogen molecule ($H_2$). These are considered epoch-making contributions in the development of *ab initio* quantum chemistry, since they provided definitive evidence for the validity of the multiple-particle Schrödinger equation for atoms and molecules. Before then, exact solutions had been obtained only for one-electron hydrogenlike atoms. The helium and hydrogen work was done long before the advent of electronic computers and required many months of drudgery, using hand-cranked calculating machines. Nowadays, students of chemistry and physics can carry out these same computations in a matter of hours, or even minutes, using relatively straightforward *Mathematica* routines. Moreover, the results can be easily improved far beyond the capabilities of the original workers.

## 1 Introduction

Neils Bohr's model of the atom, published in 1913, introduced the concept of electrons orbiting an atomic nucleus in a set of quantized energies. The characteristic line spectra of atoms could then be associated emission and absorption of electromagnetic radiation in transitions between these discrete orbits. With later elaborations by Sommerfeld and Stoner, Bohr recognized that sequential filling of the allowed electron orbits could qualitatively account for the periodic structure of the elements. However the Bohr model was *quantitatively* successful only for one-electron systems—the hydrogen atom and hydrogenlike ions such as $He^+$, $Li^{2+}$, etc. Attempts to generalize the dynamics of electron orbits failed miserably even for the two-electron helium atom. Attempted applications to molecules, with more than one nucleus, were even worse.

Classical concepts such as electron orbits were superseded by the development of quantum mechanics by Heisenberg, Schrödinger and Dirac in 1925-26. The form of quantum mechanics most suitable for the treatment of atoms and molecule was *wave mechanics*, based on the Schrödinger equation, a partial differential eigenvalue equation in the symbolic form

$$H\psi = E\psi \quad (1)$$

which could be explicitly written down for any atom or molecule. Here $H$ represents the Hamiltonian operator, $E$ is an allowed energy of the quantum system and $\psi$ is the wavefunction, a more realistic reprresentation of electrons as wavelike entities rather than orbiting particles. Schrödinger, in his original 1926 paper gave an exact solution for the hydrogen atom, in numerical agreement with Bohr's results for the allowed energy levels. The next step was to discover whether the Schrödinger equation could provide a correct description for multiple-electron atoms and for molecules. It became evident very soon that exact analytic solutions for the Schrödinger equation could *not* be obtained for any system more complicated than hydrogen atom. However, the Rayleigh-Ritz variational method could be applied to obtain *approximate* solutions to the Schrödinger equation in conformity with the inequality

$$E_0 \leq \frac{\int \psi^* H \psi \, d\tau}{\int |\psi|^2 \, d\tau} \quad (2)$$

where $E_0$ is the exact ground-state energy of the system. The game now evolved into constructing approximate wavefunctions which approched the experimental values of $E_0$, most commonly determined from spectroscopic data. An up-to-date account of the relevant quantum-mechanical background is given my recent text[1].

## 2 Helium Atom

The wavefunction $\psi$ for the $^1$S ground state of helium atom depends on three coordinates, $r_1$, $r_2$ and $r_{12}$, which form a triangle. Total orbital angular momentum of zero implies that the energy is independent of the absolute orientation of the triangle. The Schrödinger equation is given by

$$H\psi(r_1, r_2, r_{12}) = E\,\psi(r_1, r_2, r_{12}) \qquad (3)$$

with the Hamiltonian

$$H = -\frac{1}{2}\nabla_1^2 - \frac{1}{2}\nabla_2^2 - \frac{Z}{r_1} - \frac{Z}{r_2} + \frac{1}{r_{12}} \qquad (4)$$

in terms of atomic units, $\hbar = m_e = e = 1$. An infinite nuclear mass is assumed and relativistic and radiative corrections are neglected. For helium atom, the nuclear charge $Z = 2$. The energy is expressed in hartrees: 1 hartree = 27.211 electron volts. The first ionization energy for helium is experimentally 24.59 eV, while the second ionization energy is 54.42 eV. The last result is implied by the exact energy of the hydrogenlike He$^+$ ion, equal to $= -Z^2/2$ hartrees or $-54.42$ eV. Therefore the experimental ground-state energy of helium atom is given by $E_0 = -79.02$ eV $=-2.90372$ hartrees. The object now is to reproduce this value, as closely as possible, by theoretical analysis.

In the most elementary picture, the ground state of the helium atom can be described as a $1s^2$ electron configuration, meaning that both electrons occupy hydrogenlike $1s$ atomic orbitals. The first approximation to the wavefunction is therefore given by

$$\psi(r_1, r_2) \approx e^{-Z(r_1+r_2)} \qquad \text{with} \qquad Z = 2 \qquad (5)$$

The variational principle Eq (2) then given an approximate energy of $-2.75$ hartrees, certainly in the right ball park but quantitatively inferior. A fairly easy improvement is to replace the actual nuclear charge $Z$ by an effective value $\zeta$, which minimizes the variational integral. The optimal result is obtained with $\zeta = 27/16 = 1.6875$, giving a much improved ground-state energy of $-2.84765$, within about 2% of the experimental value, but still not an airtight proof of concept. The value of $\zeta < Z$ can be attributed to the partial shielding of the nuclear charge experienced by each electron, caused by the presence of the other electron.

This is where the Norwegian physicist E. A. Hylleraas[2, 3] enters the picture. Hylleraas defined new independent variables

$$s = r_1 + r_2, \qquad t = r_1 - r_2, \qquad u = r_{12} \tag{6}$$

The wavefunction is then approximated as a linear combination containing integer powers of $s$, $t$ and $u$ which we can write

$$\psi(s,t,u) = e^{-\zeta s} \sum_n c_n \, s^{\alpha_n} \, t^{\beta_n} \, u^{\gamma_n} = \sum_n c_n \, f_n(s,t,u) \tag{7}$$

We consider a sum of 10 basis functions (one more than Hylleraas):

$$f_{\{1,2\ldots10\}} = e^{-\zeta s} \times \{1,\, u,\, t^2,\, s,\, s^2,\, u^2,\, su,\, t^2 u,\, u^3,\, t^2 u^2\} \tag{8}$$

The coefficients $c_n$ are determined by the linear variational method such as to minimize the ground state energy. This involves solving the secular equation:

$$\begin{vmatrix} H_{11} - \varepsilon\, S_{11} & H_{12} - \varepsilon\, S_{12} & \cdots & H_{1N} - \varepsilon\, S_{1N} \\ H_{21} - \varepsilon\, S_{21} & H_{22} - \varepsilon\, S_{22} & \cdots & H_{2N} - \varepsilon\, S_{2N} \\ \cdots & \cdots & & \cdots \\ H_{N1} - \varepsilon\, S_{N1} & H_{N2} - \varepsilon\, S_{N2} & \cdots & H_{NN} - \varepsilon\, S_{NN} \end{vmatrix} = 0 \tag{9}$$

giving $N$ roots $\varepsilon$ representing the energy eigenvalues. We are interested only in the lowest-energy root, corresponding to the ground state. Note that the coefficients $c_n$ need not be explicitly calculated.

The matrix elements in Hylleraas coordinates are given by

$$\begin{aligned}
H_{nm} = \int_0^\infty ds \int_0^s du \int_0^u dt \, \Bigg\{ & u(s^2 - t^2)\left[\left(\frac{\partial f_n}{\partial s}\right)\left(\frac{\partial f_m}{\partial s}\right) + \left(\frac{\partial f_n}{\partial t}\right)\left(\frac{\partial f_m}{\partial t}\right) + \\
& \left(\frac{\partial f_n}{\partial u}\right)\left(\frac{\partial f_m}{\partial u}\right)\right] + s(u^2 - t^2)\left[\left(\frac{\partial f_n}{\partial u}\right)\left(\frac{\partial f_m}{\partial s}\right) + \left(\frac{\partial f_n}{\partial s}\right)\left(\frac{\partial f_m}{\partial u}\right)\right] + \\
& t(s^2 - u^2)\left[\left(\frac{\partial f_n}{\partial u}\right)\left(\frac{\partial f_m}{\partial t}\right) + \left(\frac{\partial f_n}{\partial t}\right)\left(\frac{\partial f_m}{\partial u}\right)\right] + (s^2 - t^2 - 4Zsu) f_n f_m \Bigg\}
\end{aligned} \tag{10}$$

and
$$S_{nm} = \int_0^\infty ds \int_0^s du \int_0^u dt\, u(s^2 - t^2)\, f_n f_m \qquad (11)$$

Once the matrix elements and the secular determinant are defined, the lowest root can be found in about 30 seconds using *Mathematica* 5.0 on a dual-processor Macintosh G5. The optimal value $\zeta = 1.75$ can be found from a few trial runs. The *Mathematica* notebook for the computation is appended.

The helium ground-state energy is computed to be $\varepsilon = -2.90360$ hartrees, in essential agreement with the best experimental value at the time. Subsequent theoretical work has determined that a more accurate wavefunction must be augmented by terms containing negative powers of $s$ and augmented by factors of $\ln s$. The definitive computation by Pekaris[4], using a 1078-term recursion formula, gave a nonrelativistic ground-state energy $\varepsilon = -2.903724375$ hartrees,

# 3  Hydrogen Molecule

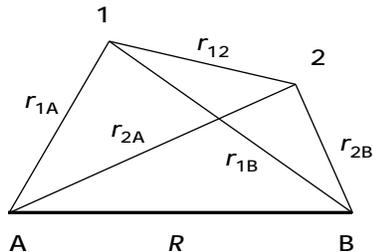

Figure 1: Coordinates in hydrogen molecule Schrödinger equation. A and B label the nuclei, 1 and 2, the electrons.

In 1933, Coolidge and James[5, 6] at Harvard carried out the first definitive *ab initio* computation on the H$_2$ molecule. The symmetry of a diatomic molecule can be exploited by using prolate spheroidal coodinates $\lambda$, $\mu$, $\varphi$, shown in Fig. 1. The Hamiltonian for the hydrogen molecule is given by

$$H = -\frac{1}{2}\nabla_1^2 - \frac{1}{2}\nabla_2^2 - \frac{1}{r_{1a}} - \frac{1}{r_{1b}} - \frac{1}{r_{2a}} - \frac{1}{r_{2b}} + \frac{1}{r_{12}} + \frac{1}{R} \qquad (12)$$

where the variables are shown in Fig. 2. In accordance with the Born-Oppenheimer approximation, the internuclear distance $R$ is assumed fixed in computations of electronic energy. For $\Sigma$ states of the molecule, orbital angular momentum components along the internuclear axis vanish, and the wavefunction is cylindrically-symmetrical. Wavefunctions can be constructed as functions of five dimensionless variables:

$$\lambda_1 = \frac{r_{1a} + r_{1b}}{R}, \quad \lambda_2 = \frac{r_{2a} + r_{2b}}{R}, \quad \mu_1 = \frac{r_{1a} - r_{1b}}{R}, \quad \mu_2 = \frac{r_{2a} - r_{2b}}{R} \quad (13)$$

and

$$\rho = \frac{2r_{12}}{R} \quad (14)$$

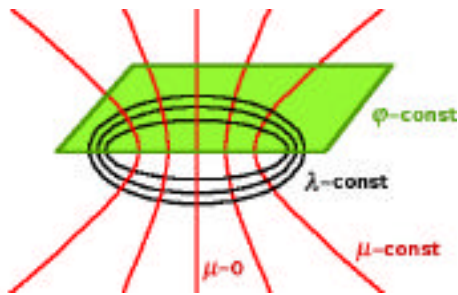

Figure 2: Prolate spheroidal coordinates.

James and Coolidge considered variational functions of the form

$$\psi(\lambda_1, \lambda_2, \mu_1, \mu_2, \rho) = \sum_{mnjkp} c_{mnjkp} f_{mnjkp} \quad (15)$$

with

$$f_{mnjkp} = \frac{1}{2\pi} e^{-\alpha(\lambda_1 + \lambda_2)} \left( \lambda_1^m \lambda_2^n \mu_1^j \mu_2^k \rho^p + \lambda_1^n \lambda_2^m \mu_1^k \mu_2^j \rho^p \right) \quad (16)$$

where $m, n, j, k$ are integers $\geq 0$, $p$ is an integer $\geq -1$ and $j + k$ must be even. James and Coolidge used a sum of 13 basis functions. *Mathematica* finds essentially the same result with 11 selected functions. (It is difficult to solve a determinantal equation larger than $11 \times 11$). The computation takes approximately 10 hours (James and Coolidge took the better part of a year). The optimized molecular energy of $-1.17300$ hartrees is obtained with the exponential parameter $\alpha = 1.5$ at the equilibrium internuclear distance $R = 1.40$ bohr. Our computation gives a molecular energy of $-1.17300$ hartree at the equilibrium $R$. Since the energy of two separated hydrogen atoms equals $-1$ hartree, the binding energy $D_e = 4.7075$ eV. Subsequent computationskw using up to 100-term expansions gives values of $\varepsilon = -1.17447$ hartrees and $D_e = 4.7475$ eV.

It is sufficient for present purposes to describe a scaled-down version of the James-Coolidge computation, using just 5 terms, which takes about 70 minutes to run. A molecular energy of $-1.6647$ and binding energy of $D_e = 4.53$ eV are obtained.

The interelectronic variable $\rho$ can be expressed in terms of the prolate spheroidal coordinates using

$$\rho^2 = \lambda_1^2 + \lambda_2^2 + \mu_1^2 + \mu_2^2 - 2 - 2\lambda_1\lambda_2\mu_1\mu_2$$

$$-2\left[(\lambda_1^2 - 1)(\lambda_2^2 - 1)(1 - \mu_1^2)(1 - \mu_2^2)\right]^{1/2} \cos(\varphi_1 - \varphi_2) \quad (17)$$

and its reciprocal given by a Neumann expansion:

$$\frac{1}{\rho} = \sum_{\tau=0}^{\infty} \sum_{\nu=0}^{\tau} \epsilon_\nu \, P_\tau^\nu(\lambda_<) Q_\tau^\nu(\lambda_>) P_\tau^\nu(\mu_1) P_\tau^\nu(\mu_2) \cos[\nu(\varphi_1 - \varphi_2)] \quad (18)$$

where $\epsilon_0 = 2\tau + 1$ and $\epsilon_{\nu>0} = 2(-1)^\nu(2\tau + 1)[(\tau - \nu)!/(\tau + \nu)!]^2$. the variables $\lambda_>, \lambda_<$ are the greater and lesser of $\lambda_1, \lambda_2$, while $P_\tau^\nu$ $Q_\tau^\nu$ are associated Legendre function of the first and second kind, respectively. Since $-1 \leq \mu \leq +1$, $P_\tau^\nu(\mu)$ needs the *Mathematica* subroutine for LegendreP$[\tau, \nu, 1, \mu]$, while $P_\tau^\nu(\lambda)$ and $Q_\tau^\nu(\lambda)$, with the ranges $1 \leq \lambda \leq \infty$, require LegendreP$[\tau, \nu, 3, \lambda]$ and LegendreQ$[\tau, \nu, 3, \lambda]$, respectively. Terms linear in $\rho$ are found from products of (17) and (18).

The matrix elements of $H$ and $S$ can be expressed in terms of the 6-fold integrals:

$$X_{mnjkp} = \frac{1}{4\pi^2} \int \cdots \int d\lambda_1 \, d\lambda_2 \, d\mu_1 \, d\mu_2 \, \partial\varphi_1 \, \partial\varphi_2 (\lambda_1^2 - \mu_1^2) e^{-\alpha(\lambda_1+\lambda_2)} \lambda_1^m \lambda_2^n \mu_1^j \mu_2^k \rho^p \quad (19)$$

The requisite formulas are fairly lengthy and we leave the details to the appended *Mathematica* printout.

```
(* Hylleraas Computation on Helium Atom *)

f[1] := e^(-ζ s)

f[2] := e^(-ζ s) u

f[3] := e^(-ζ s) t^2

f[4] := e^(-ζ s) s

f[5] := e^(-ζ s) s^2

f[6] := e^(-ζ s) u^2

f[7] := e^(-ζ s) s u

f[8] := e^(-ζ s) t^2 u

f[9] := e^(-ζ s) u^3

f[10] := e^(-ζ s) t^2 u^2
```

$$S[n\_, m\_] := \int_0^\infty \left( \int_0^s \left( \int_0^u f[n]\, f[m]\, u\, (s^2 - t^2)\, dt \right) du \right) ds$$

$$k1[n\_, m\_] := ((\partial_s f[n])(\partial_s f[m]) + (\partial_t f[n])(\partial_t f[m]) + (\partial_u f[n])(\partial_u f[m]))\, u\, (s^2 - t^2)$$

$$K1[n\_, m\_] := \int_0^\infty \left( \int_0^s \left( \int_0^u k1[n, m]\, dt \right) du \right) ds$$

$$k2[n\_, m\_] := ((\partial_s f[n])(\partial_u f[m]) + (\partial_u f[n])(\partial_s f[m]))\, s\, (u^2 - t^2) + ((\partial_t f[n])(\partial_u f[m]) + (\partial_u f[n])(\partial_t f[m]))\, t\, (s^2 - u^2)$$

$$K2[n\_, m\_] := \int_0^\infty \left( \int_0^s \left( \int_0^u k2[n, m]\, dt \right) du \right) ds$$

$$Z = 2;\quad v[n\_, m\_] := f[n]\, f[m]\, (s^2 - t^2 - 4\, Z\, s\, u)$$

$$V[n\_, m\_] := \int_0^\infty \left( \int_0^s \left( \int_0^u v[n, m]\, dt \right) du \right) ds$$

```
H[n_, m_] := K1[n, m] + K2[n, m] + V[n, m]

M[n_, m_] := H[m, n] - ϵ S[m, n]

ζ = 1.75;

Timing[NSolve[Det[Array[M, {10, 10}]] == 0, ϵ]]
```

{39.04 Second, {{ϵ → -2.9036}, {ϵ → -1.97036}, {ϵ → -1.17629}, {ϵ → -0.367166}, {ϵ → 0.575448}, {ϵ → 0.790687}, {ϵ → 2.51695}, {ϵ → 5.63955}, {ϵ → 7.16528}, {ϵ → 11.0266}}}

**Ground state : ϵ = -2.903602 hartree**

```
(* James & Coolidge 5-Term Computation on Hydrogen Molecule  *)

(* negative indices refer to interchanged {m,n}, {j,k} *)

m[x_] := 0; m[3] = 1;

n[x_] := 0; n[-3] = 1;

j[x_] := 0; j[4] = 1; j[-4] = 1; j[-2] = 2;

k[x_] := 0; k[4] = 1; k[2] = 2; k[-4] = 1;

p[x_] := 0; p[5] = p[-5] = 1;

Do[Print[{m[r], n[r], j[r], k[r], p[r]},
   "     ", {m[-r], n[-r], j[-r], k[-r], p[-r]}], {r, 1, 5}]
```

{0, 0, 0, 0, 0}     {0, 0, 0, 0, 0}

{0, 0, 0, 2, 0}     {0, 0, 2, 0, 0}

{1, 0, 0, 0, 0}     {0, 1, 0, 0, 0}

{0, 0, 1, 1, 0}     {0, 0, 1, 1, 0}

{0, 0, 0, 0, 1}     {0, 0, 0, 0, 1}

```
α[a_, b_] := m[a] + m[b]; α1[a_, b_] := m[a] - m[b];
β[a_, b_] := n[a] + n[b]; β1[a_, b_] := n[a] - n[b];
γ[a_, b_] := j[a] + j[b]; γ1[a_, b_] := j[a] - j[b]; δ[a_, b_] := k[a] + k[b];
δ1[a_, b_] := k[a] - k[b]; ϵ[a_, b_] := p[a] + p[b]; ϵ1[a_, b_] := p[a] - p[b]
```

General::spell1 : Possible spelling error: new symbol name "β1" is similar to existing symbol "α1". More…

General::spell : Possible spelling error: new symbol name "γ1" is similar to existing symbols {α1, β1}. More…

General::spell :
 Possible spelling error: new symbol name "δ1" is similar to existing symbols {α1, β1, γ1}. More…

General::spell :
 Possible spelling error: new symbol name "ϵ1" is similar to existing symbols {α1, β1, γ1, δ1}. More…

General::stop : Further output of General::spell will be suppressed during this calculation. More…

**(* Ignore warning, we very carefully defined each symbol! *)**

**(* General formula: $X[\alpha\_,\beta\_,\gamma\_,\delta\_,\epsilon\_] := \frac{1}{4\pi^2}$**
$\int_1^\infty \left(\int_1^\infty \left(\int_{-1}^1 \left(\int_{-1}^1 \left(\int_0^{2\pi} \left(\int_0^{2\pi} ((\lambda 1^2 - \mu 1^2)\, e^{-1.5\,(\lambda 1 + \lambda 2)}\, \lambda 1^\alpha\, \lambda 2^\beta\, \mu 1^\gamma\, \mu 2^\delta\, \rho^\epsilon\right)\, d\varphi 2\right) d\varphi 1\right) d\mu 2\right) d\mu 1\right) d\lambda 2\right) d\lambda 1$ *)

$$X[a\_, b\_, c\_, d\_, 0] := \int_1^\infty \left( \int_1^\infty \left( \int_{-1}^1 \left( \int_{-1}^1 ((\lambda 1^2 - \mu 1^2) \, e^{-1.5(\lambda 1 + \lambda 2)} \, \lambda 1^a \, \lambda 2^b \, \mu 1^c \, \mu 2^d) \, d\mu 2 \right) d\mu 1 \right) d\lambda 2 \right) d\lambda 1$$

General::spell :
 Possible spelling error: new symbol name "λ1" is similar to existing symbols {α1, β1, γ1, δ1, ϵ1}. More…

General::spell :
 Possible spelling error: new symbol name "μ1" is similar to existing symbols {α1, β1, γ1, δ1, ϵ1, λ1}. More…

General::spell1 : Possible spelling error: new symbol name "μ2" is similar to existing symbol "λ2". More…

$$X[a\_, b\_, c\_, d\_, 2] :=$$
$$\int_1^\infty \left( \int_1^\infty \left( \int_{-1}^1 \left( \int_{-1}^1 ((\lambda 1^2 - \mu 1^2)(\lambda 1^2 + \lambda 2^2 + \mu 1^2 + \mu 2^2 - 2 - 2\lambda 1\lambda 2\mu 1\mu 2) \, e^{-1.5(\lambda 1+\lambda 2)} \, \lambda 1^a \, \lambda 2^b \, \mu 1^c \, \mu 2^d) \right.\right.\right.$$
$$\left.\left.\left. d\mu 2 \right) d\mu 1 \right) d\lambda 2 \right) d\lambda 1$$

$$P0[\tau\_, \nu\_, z\_] := \text{LegendreP}[\tau, \nu, 1, z]$$

$$P[\tau\_, \nu\_, z\_] := \text{LegendreP}[\tau, \nu, 3, z]$$

$$Q[\tau\_, \nu\_, z\_] := \text{LegendreQ}[\tau, \nu, 3, z]$$

$$X[a\_, b\_, c\_, d\_, -1] :=$$
$$\sum_{\tau=0}^{\text{Min}[c+2,d]} (2\tau+1) \left( \text{NIntegrate}\left[ \int_{-1}^1 \left( \int_{-1}^1 ((\lambda 1^2 - \mu 1^2) \, e^{-1.5(\lambda 1+\lambda 2)} \, P[\tau, 0, \lambda 1] \, Q[\tau, 0, \lambda 2] \, P0[\tau, 0, \mu 1] \right.\right.\right.$$
$$\left.\left. P0[\tau, 0, \mu 2] \, \lambda 1^a \, \lambda 2^b \, \mu 1^c \, \mu 2^d) \, d\mu 2 \right) d\mu 1, \{\lambda 2, 1, 15\}, \{\lambda 1, 1, \lambda 2\} \right] +$$
$$\text{NIntegrate}\left[ \int_{-1}^1 \left( \int_{-1}^1 ((\lambda 1^2 - \mu 1^2) \, e^{-1.5(\lambda 1+\lambda 2)} \, P[\tau, 0, \lambda 2] \, Q[\tau, 0, \lambda 1] \, P0[\tau, 0, \mu 1] \right.\right.$$
$$\left.\left.\left. P0[\tau, 0, \mu 2] \, \lambda 1^a \, \lambda 2^b \, \mu 1^c \, \mu 2^d) \, d\mu 2 \right) d\mu 1, \{\lambda 1, 1, 15\}, \{\lambda 2, 1, \lambda 1\} \right] \right)$$

```
X[a_, b_, c_, d_, 1] := Chop[
  Sum[(2 τ + 1) (NIntegrate[Integrate[Integrate[((λ1^2 - μ1^2) E^(-1.5 (λ1+λ2)) (λ1^2 + λ2^2 + μ1^2 + μ2^2 - 2 - 2 λ1 λ2 μ1 μ2) P[τ, 0, λ1] Q[τ, 0, λ2] P0[τ, 0, μ1] P0[τ, 0, μ2] λ1^a λ2^b μ1^c μ2^d), {μ2, -1, 1}], {μ1, -1, 1}], {λ2, 1, 15}, {λ1, 1, λ2}] + NIntegrate[Integrate[Integrate[((λ1^2 - μ1^2) E^(-1.5 (λ1+λ2)) (λ1^2 + λ2^2 + μ1^2 + μ2^2 - 2 - 2 λ1 λ2 μ1 μ2) P[τ, 0, λ2] Q[τ, 0, λ1] P0[τ, 0, μ1] P0[τ, 0, μ2] λ1^a λ2^b μ1^c μ2^d), {μ2, -1, 1}], {μ1, -1, 1}], {λ1, 1, 15}, {λ2, 1, λ1}]), {τ, 0, Min[c+4, d+2]}] +
  Sum[2 (2 τ + 1) ((τ-1)!/(τ+1)!)^2 (NIntegrate[Integrate[Integrate[((λ1^2 - μ1^2) E^(-1.5 (λ1+λ2)) Sqrt[(λ1^2 - 1) (λ2^2 - 1) (1 - μ1^2) (1 - μ2^2)] P[τ, 1, λ1] Q[τ, 1, λ2] P0[τ, 1, μ1] P0[τ, 1, μ2] λ1^a λ2^b μ1^c μ2^d), {μ2, -1, 1}], {μ1, -1, 1}], {λ2, 1, 15}, {λ1, 1, λ2}] +
      NIntegrate[Integrate[Integrate[((λ1^2 - μ1^2) E^(-1.5 (λ1+λ2)) Sqrt[(λ1^2 - 1) (λ2^2 - 1) (1 - μ1^2) (1 - μ2^2)] P[τ, 1, λ2] Q[τ, 1, λ1] P0[τ, 1, μ1] P0[τ, 1, μ2] λ1^a λ2^b μ1^c μ2^d), {μ2, -1, 1}], {μ1, -1, 1}], {λ1, 1, 15}, {λ2, 1, λ1}]), {τ, 1, Min[c+3, d+1]}]]

R = 1.40;

s[a_, b_] := R^6/64 (X[α[a, b], β[a, b] + 2, γ[a, b], δ[a, b], ϵ[a, b]] -
    X[α[a, b], β[a, b], γ[a, b], δ[a, b] + 2, ϵ[a, b]])

S[a_, b_] := s[a, b] + s[-a, b] + s[a, -b] + s[-a, -b]

h1[a_, b_] :=
  -R (X[α[a, b], β[a, b] + 2, γ[a, b], δ[a, b], ϵ[a, b]] - X[α[a, b], β[a, b], γ[a, b],
      δ[a, b] + 2, ϵ[a, b]] + 2 X[α[a, b], β[a, b] + 2, γ[a, b], δ[a, b], ϵ[a, b] - 1] -
    2 X[α[a, b], β[a, b], γ[a, b], δ[a, b] + 2, ϵ[a, b] - 1] -
    8 X[α[a, b], β[a, b] + 1, γ[a, b], δ[a, b], ϵ[a, b]])

h2[a_, b_] := (β1[a, b]^2 - δ1[a, b]^2 + β[a, b] - δ[a, b] + ϵ1[a, b] (β1[a, b] - δ1[a, b]))
  X[α[a, b], β[a, b], γ[a, b], δ[a, b], ϵ[a, b]]

h3[a_, b_] := -2 * 1.5 X[α[a, b], β[a, b] + 1, γ[a, b], δ[a, b], ϵ[a, b]]

h4[a_, b_] := -(β1[a, b]^2 - β[a, b]) *
  If[β1[a, b]^2 - β[a, b] == 0, 0, X[α[a, b], β[a, b] - 2, γ[a, b], δ[a, b], ϵ[a, b]]]

h5[a_, b_] := (δ1[a, b]^2 - δ[a, b]) *
  If[δ1[a, b]^2 - δ[a, b] == 0, 0, X[α[a, b], β[a, b], γ[a, b], δ[a, b] - 2, ϵ[a, b]]]
```

```
h6[a_, b_] := (ϵ1[a, b]^2 + ϵ[a, b] + ϵ1[a, b] (β1[a, b] + δ1[a, b])) *
   If[(ϵ1[a, b]^2 + ϵ[a, b] + ϵ1[a, b] (β1[a, b] + δ1[a, b])) == 0,
    0, (X[α[a, b], β[a, b] + 2, γ[a, b], δ[a, b], ϵ[a, b] - 2] -
      X[α[a, b], β[a, b], γ[a, b], δ[a, b] + 2, ϵ[a, b] - 2])]

h7[a_, b_] := -ϵ1[a, b] (β1[a, b] - δ1[a, b]) *
   If[ϵ1[a, b] (β1[a, b] - δ1[a, b]) == 0, 0, (X[α[a, b] + 2, β[a, b], γ[a, b],
      δ[a, b], ϵ[a, b] - 2] + X[α[a, b], β[a, b], γ[a, b] + 2, δ[a, b], ϵ[a, b] - 2])]

h8[a_, b_] := 2 ϵ1[a, b] β1[a, b] *
   If[ϵ1[a, b] β1[a, b] == 0, 0, X[α[a, b] + 1, β[a, b] - 1, γ[a, b] + 1, δ[a, b] + 1, ϵ[a, b] - 2]]

h9[a_, b_] := -2 ϵ1[a, b] δ1[a, b] *
   If[ϵ1[a, b] δ1[a, b] == 0, 0, X[α[a, b] + 1, β[a, b] + 1, γ[a, b] + 1, δ[a, b] - 1, ϵ[a, b] - 2]]

h[a_, b_] := - R^4/64
   (h1[a, b] + h2[a, b] + h3[a, b] + h4[a, b] + h5[a, b] + h6[a, b] + h7[a, b] + h8[a, b] + h9[a, b])

H[a_, b_] := h[a, b] + h[-a, b] + h[a, -b] + h[-a, -b]

M[a_, b_] := H[a, b] - λ S[a, b]

Timing[NSolve[Det[Array[M, {5, 5}]] == 0, λ]]
```

NIntegrate::ploss :
 Numerical integration stopping due to loss of precision. Achieved neither the requested PrecisionGoal nor AccuracyGoal; suspect one of the following: highly oscillatory integrand or the true value of the integral is 0. If your integrand is oscillatory try using the option Method->Oscillatory in NIntegrate. More…

NIntegrate::ploss :
 Numerical integration stopping due to loss of precision. Achieved neither the requested PrecisionGoal nor AccuracyGoal; suspect one of the following: highly oscillatory integrand or the true value of the integral is 0. If your integrand is oscillatory try using the option Method->Oscillatory in NIntegrate. More…

NIntegrate::ploss :
 Numerical integration stopping due to loss of precision. Achieved neither the requested PrecisionGoal nor AccuracyGoal; suspect one of the following: highly oscillatory integrand or the true value of the integral is 0. If your integrand is oscillatory try using the option Method->Oscillatory in NIntegrate. More…

General::stop : Further output of NIntegrate::ploss will be suppressed during this calculation. More…

{4172.16 Second,
 {{λ → -1.16647}, {λ → -0.336828}, {λ → 0.223073}, {λ → 0.841628}, {λ → 1.24718}}}

**.1664719858005372 * 27.211 eV**

4.52987 eV